\documentclass{ws-procs9x6}
\usepackage{epsfig}
\usepackage{wrapfig}

\begin{document}

% A useful Journal macro
\def\Journal#1#2#3#4{{#1} {\bf #2}, #3 (#4)}
\def\CPC{\em Comp. Phys. Comm.~}
\def\NCA{\em Nuovo Cimento~}
\def\NIM{\em Nucl. Instrum. Methods~}
\def\NIMA{{\em Nucl. Instrum. Methods} A}
\def\NPB{{\em Nucl. Phys.} B}
\def\PLB{{\em Phys. Lett.}  B}
\def\PRL{\em Phys. Rev. Lett.~}
\def\PRD{{\em Phys. Rev.} D}
\def\ZPC{{\em Z. Phys.} C}
\def\EPC{{\em Eur. Phys. J.} C}

\newcommand{\pom}{I\!\!P}
\newcommand{\reg}{I\!\!R}
\newcommand{\alphapom}{\alpha_{_{I\!\!P}}}
\newcommand{\alphareg}{\alpha_{_{\rm I\!R}}}
\newcommand{\xpom}{x_{\pom}}
\newcommand{\xPom}{\xpom}
\newcommand{\xbj}{$x_{Bj}$} 
\newcommand{\qsq}{\mbox{$Q^2$}}
\newcommand{\bet} {$\beta$}
\newcommand{\wsq} {$W^2$}
\newcommand{\ttt} {$t$}
\newcommand{\fdfourf} {$F_2^{D(4)}  (Q^2, x_{I\!\!P} , \beta , t)$}
\newcommand{\fdfour} {$F_2^{D(4)}$}
\newcommand{\fdthree} {$F_2^{D(3)}$}
\newcommand{\fdthreef} {$F_2^{D(3)}  (Q^2, x_{I\!\!P} , \beta )$}
\newcommand{\fdtwo} {$F_2^{D(2)}$}
\newcommand{\fdtwof} {$F_2^{D(2)} (Q^2,\beta)$}
\newcommand{\ftwo} {$F_2 (Q^2,x)$}

\title{Inclusive diffraction\footnote{Contribution to the Proceedings
of the Ringberg Workshop ``New Trends in HERA Physics 2005"}}

\author{L.~FAVART}

\address{I.I.H.E., CP-230\\
        Universit\'e Libre de Bruxelles, \\
        1050 Brussels\\
        Belgium\\
        lfavart@ulb.ac.be}

\maketitle

\abstracts{
Results are given on the measurements of the hard
diffractive interactions at HERA $ep$ collider.
The structure of the diffractive exchange in terms of partons and the
factorisation properties are discussed, in particular by comparing the 
QCD predictions for dijets and $D^*$ with measurements both in the 
photo and electroproduction regimes.
}

\section{Introduction}

%The diffractive interaction is a feature of hadron-hadron scattering
%at high energy corresponding to a $t$-channel exchange of the vacuum
%quantum numbers and a small momentum transfer.
%It has been described, in the past, in the framework of Regge 
%theory, where the exchange was interpreted as the Pomeron ($\pom$) trajectory, 
%characterized by a weak energy dependence, in particular, with respect to
%the fast decrease of the total cross section at smaller energy due to
%Reggeon ($\reg$) exchange:
%$\sigma_{Tot}=B\ W^{2(\alphareg -1)} + \ A \ W^{2(\alphapom -1)}$,
%where $W$ is the center of mass energy, $A$ and $B$ normalisation
%factors and typically $\alphareg=0.55$ and $\alphapom=1.08$.
%\\
%

%Diffraction, directly related to the total cross section through the
%optical theorem, has two distinguishing features. 
%First, hadron emission from the exchange itself being suppressed
%by its colourless nature, the two diffractively dissociated
%systems are separated in rapidity space, forming
%a large rapidity gap (LRG).
%Secondly, the diffractive events are observed with a small momentum 
%transfer in both the transverse and longitudinal coordinates.
%The four-momentum of the exchange, $t$,
%and the longitudinal
%momentum fraction of the exchange, $\xpom$, are both small;
%$|t|$ is typically less than the square of the nucleon
%mass and $\xpom$ is smaller than 0.05.
%\\

\begin{figure}[thbp]
 %\vspace*{0.5cm}
 \begin{picture}(60,80) 
  \put(30,-2){\epsfig{figure=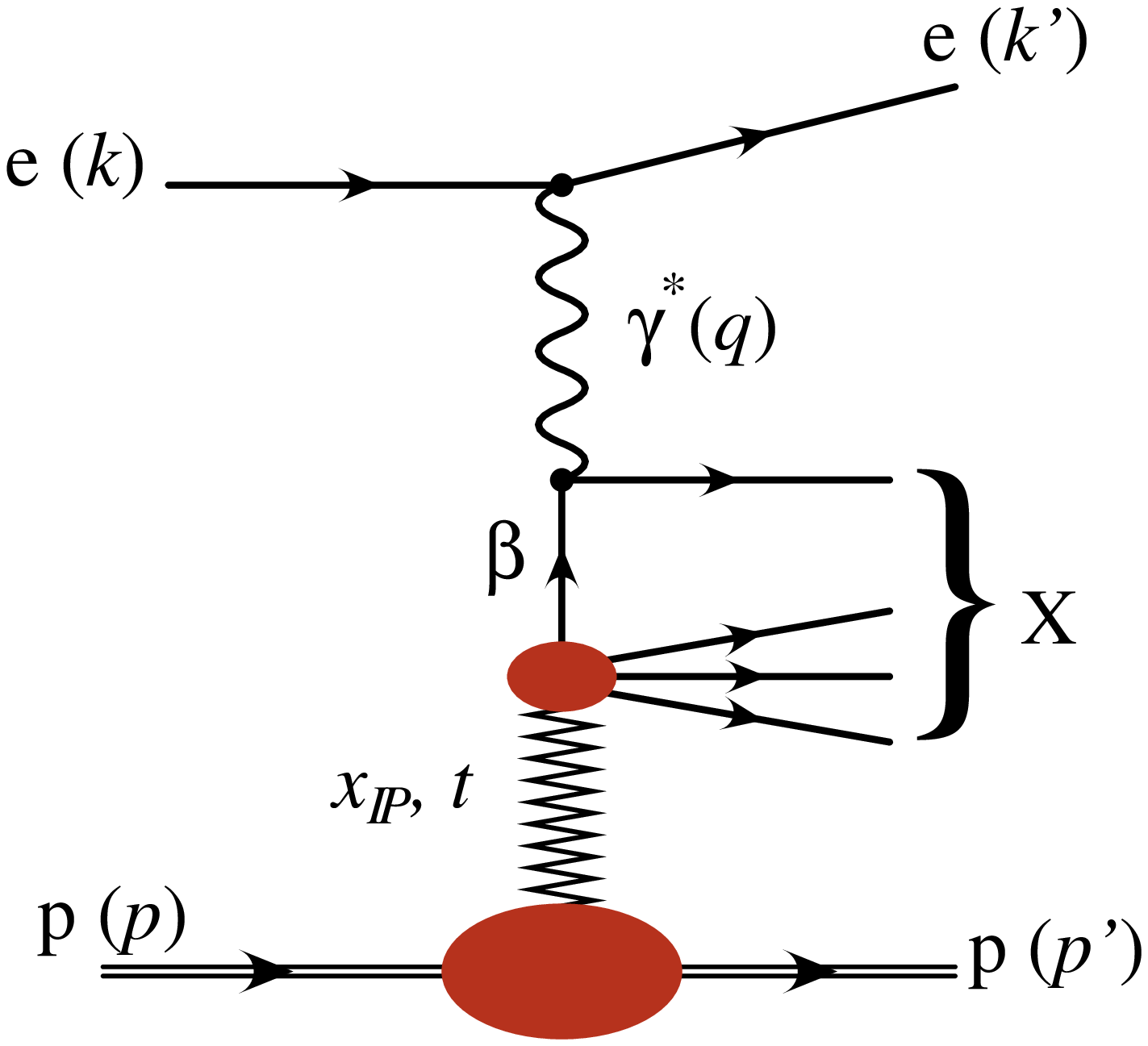,width=0.30\textwidth}} 
  \put(200,-2){\epsfig{figure=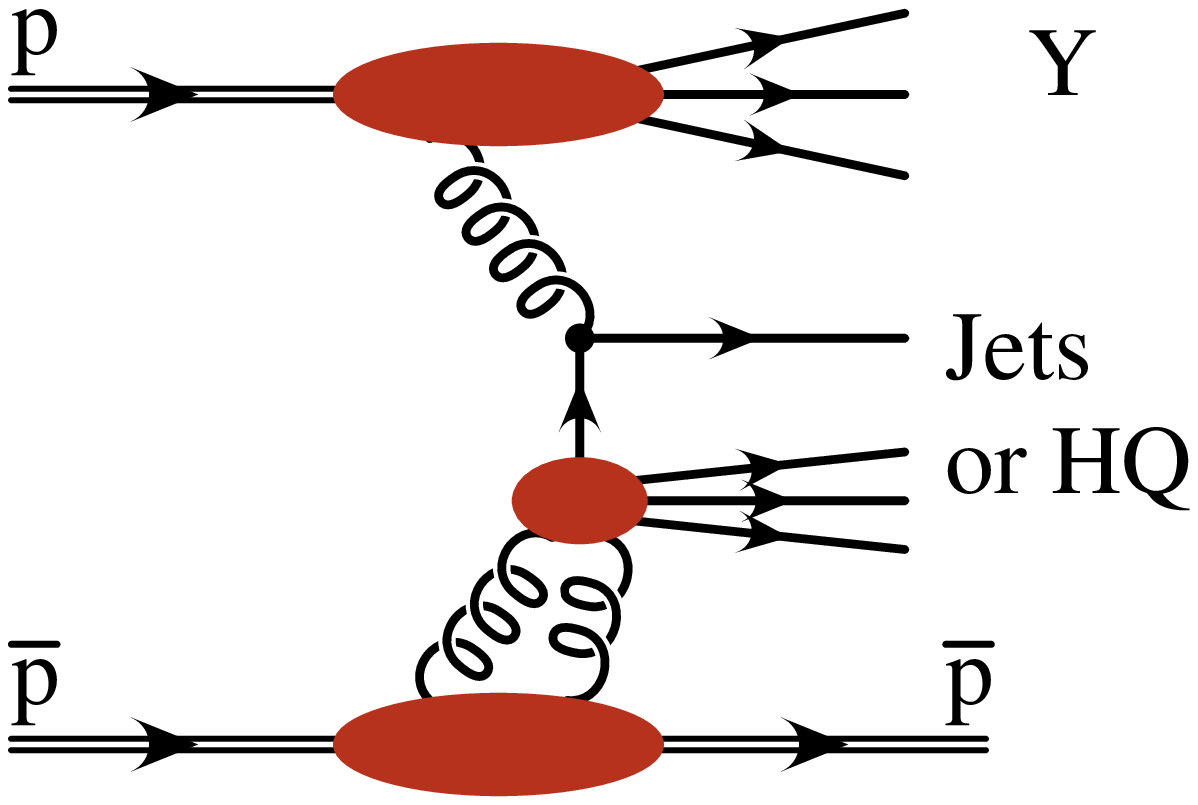,width=0.35\textwidth}} 
  \put(35,80){a) HERA}
  \put(220,80){b) Tevatron}
 \end{picture}
 \caption{Basic diagrams for diffraction in presence of a hard scale at
 HERA and at Tevatron}
 \label{fig:diag}
\end{figure}

The high energies of the HERA $ep$ collider and of the Tevatron
$p\bar{p}$ collider allow us for the first time to study diffraction 
in terms of perturbative QCD (pQCD), i.e.~in the presence of a hard
scale, which is called hard diffraction. 
At HERA the diffractive interaction takes place between the
hadronic behaviour of the exchanged virtual photon and the proton
(see Fig.~\ref{fig:diag}a), and 
between the two protons at Tevatron (see Fig.~\ref{fig:diag}b). 
The possible hard scales are large $Q^2$, 
the negative of the four-momentum squared of the exchanged
virtual photon, large transverse energy, $E_T$, in jet production,
heavy-quark mass or large momentum transfer squared at the proton
vertex, $t$.
The hard scale presence of large $Q^2$ at HERA and large  $E_T$ at Tevatron,
in particular,
gives the possibility to probe the partonic structure of the diffractive
exchange. 
%As in inclusive inelastic scattering (DIS)~\cite{Vladimir}, 
%the presence of a hard scale in the process is characterized by an 
%important positive increase of the energy dependence behaviour. 
\\

This article concentrates on the factorisation properties
using a structure function approach 
giving insight into the understanding of the nature of diffraction in
terms of partons. 
Before reviewing recent measurements, a short
discussion on the parton densities of the diffractive exchange and their 
relation to the diffractive cross section is given.
\\

Although it will not be covered here, the study of diffraction has
several other topics of interest on top of the diffractive exchange measurement
in terms of partons. It gives access to the partons correlation through
the exclusive final state measurements (generalised parton distribution
formalism). 
It allows to test the region of validity
of the different asymptotic dynamical approaches of QCD that are DGLAP 
and BFKL from the measurement of vector meson or photon exclusive
production at large values of $|t|$. 
Finally, the alternative approach of colour Dipole models 
allows us to study the transition between non-perturbative and perturbative
regimes and to test the presence of a gluon density saturation in the
proton. 

\section{Partonic Structure of Diffractive Exchanges and Factorisation
Properties}

The inclusive diffractive cross section at HERA, $e p \rightarrow e X p$,
can be defined with the help of four kinematic variables
conveniently chosen as  \qsq , $\xpom$ , \bet\ and \ttt, where
$\xpom$ \ and \bet\ are defined as

$$ 
   \xpom \simeq \frac{Q^2 + M_X^2}{Q^2 + W^2 },
   \qquad \qquad 
    \beta \simeq \frac{Q^2 }{Q^2 + M_X^2} ; \label{eq:kin}
$$ 
$M_X$ being the invariant mass of the X system, and $W$ the $\gamma^*-p$
center of mass energy.
$\xpom$ can be interpreted as the fraction of the
proton momentum carried by the exchanged
Pomeron and \bet\  is the fraction of the exchanged momentum carried by
the quark struck by the photon, or in other terms, the fraction of the
exchanged momentum reaching the photon.
These variables are related to the Bjorken \xbj\ scaling variable
by the relation \xbj$= \beta \cdot \xpom$.
The presence of the hard scale, $Q^2$, ensures that the virtual photon
is point-like and that the photon probes the partonic 
structure of the diffractive exchange (Fig.~\ref{fig:diag}a),
in analogy with the inclusive DIS processes.
\\

{\bf Factorisation Properties in hard Diffraction.}\\
For hard QCD processes in general, like high-$E_T$ jet production or
DIS,
the cross section can be factorized into two terms: the parton
density and the hard parton-parton cross section. 
The cross section can be written as
$$
\sigma = \sum_i {f_i(\xi, \mu^2) \; \hat{\sigma}_{i\gamma}(\xi, \mu^2)}
$$
where $i$ runs over all parton types, $f_i$ is the parton density
function for the $i$-th parton with longitudinal momentum fraction
$\xi$, which is probed at the factorisation scale $\mu$.
$\hat{\sigma}_{i\gamma}$
denotes the cross section for the interaction of
the $i$-th parton and the virtual photon.
Such an expression, often referred to as the QCD factorisation theorem,
is well
supported by the data. If the theorem holds,
only one parton per hadron is coupled to the hard scattering vertex.
\\

The theorem is proven to be applicable at all orders in the strong force 
coupling constant $\alpha_s$ for the leading log $Q^2$ for 
hard inclusive diffraction~\cite{collins_fact} in $ep$ collisions 
at large \qsq, namely
$$
\frac{d\sigma(x,Q^2,\xPom,t)}{d\xPom dt} = 
     \sum_i\int^{\xPom}_{x} dz \;
      \hat{\sigma}_{i\gamma}(z, Q^2, \xPom) \;
             f_i^D(z, Q^2, \xPom, t)
$$
where $z$ is the longitudinal momentum fraction of the parton
in the proton, $\hat{\sigma}_{i\gamma}$ is again the hard scattering
parton-photon cross section for hard diffraction and $f_i^D$
is the diffractive parton density for the $i$-th parton.
$f_i^D$ can be regarded as the parton density of the diffractive
exchange occuring at a given $(\xPom, t)$.
If such a theorem holds, $f_i^D$ should be universal for all hard
processes,
e.g.\ inclusive diffraction, jet or heavy-quark production etc. 
\\

If the scattered proton is not detected in Roman Pots, 
the \ttt\ variable is not measured with enough accuracy and is
integrated over.
In analogy with non-diffractive DIS scattering, the measured cross
section 
is expressed, in the neutral current case, in the form of a 
three-fold diffractive structure function
\fdthreef\ (neglecting the longitudinal contribution and $Z$ exchange), 
$$
  \frac { {\rm d}^3 \sigma \ (e p \rightarrow e X p) }
 { {\rm d}Q^2 \ {\rm d}\xpom \ {\rm d}\beta}
        = \frac {4 \pi \alpha^2} {\beta Q^4}
            \ (1 - y + \frac {y^2}{2} )
            \ F_2^{D(3)} (Q^2, \xpom , \beta) ,
                                            \label{eq:fdthreefull}
$$
where $y$ is the usual scaling variable, with
$  y \simeq W^2 / s$.
\\

\begin{figure}[htb]
 \vspace*{-0.8cm}
 \begin{center}
  \epsfig{figure=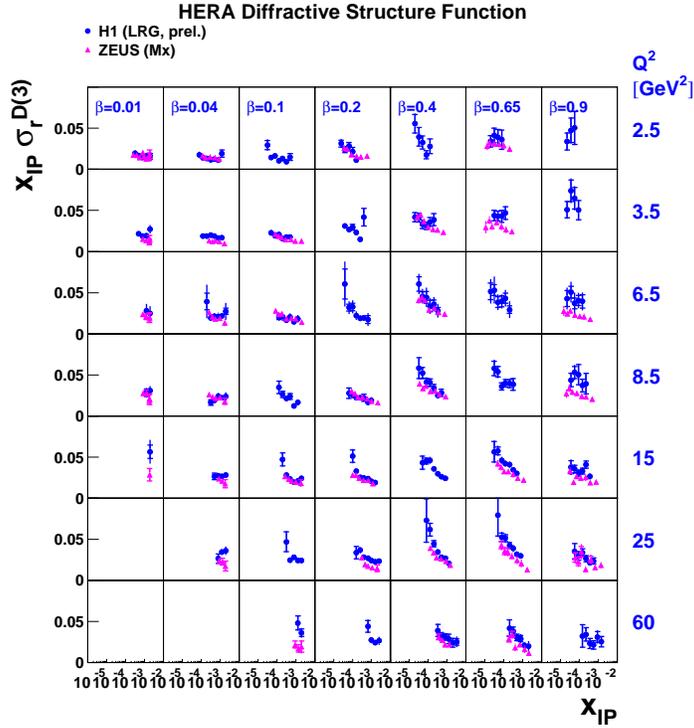,width=0.82\linewidth}
 \end{center}
 \caption{H1 and ZEUS reduced cross section (equal to \fdthree\ if
$F_L^D$ and $F_3^D$ can be neglected) 
measurements as a function $\xpom$ for fixed values of \qsq\ and \bet. 
}
%\vspace*{-0.5cm}
 \label{fig:f2d}
\end{figure}

Conveniently, the Regge factorisation is applied where \fdthree \ is 
written in the form
$$F_2^{D(3)} (Q^2, \xpom , \beta ) =
     f_{\pom/p} (\xpom ) \cdot F_2^D (Q^2, \beta),$$
using the approximation that the pomeron ($\pom$) flux $ f_{\pom/p} (\xpom)$ 
is independent
of the  $\pom$ structure \fdtwo,
by analogy with the hadron structure functions, \bet\ playing the role
of Bjorken \xbj. The  $\pom$ flux is parameterised using a Regge inspired
form $f_{\pom/p} = e^{bt}/\xpom^{2\alphapom(t) - 1}$.
\\

 H1~\cite{h1f2d97} and ZEUS \cite{zeuscharm} measurements 
of \fdthree\ diffractive structure function
are presented in Fig.\ \ref{fig:f2d}
as a function of $\xpom$ for fixed values of \qsq\ and \bet. 
The reduced cross section $\sigma_r^D$ presented in the figure 
is equal  
to \fdthree\ if $F_L^D$ and $F_3^D$ can be neglected. 

Both measurements are in good general agreement, although the \qsq\
dependence at fixed \bet\ is stronger in the H1 measurement and the last
bin in \bet\ exhibits higher values for the H1 measurement for 
$Q^2<25$ GeV$^2$.
\\

To fit the \fdthreef\ points H1 includes a sub-leading Reggeon ($\reg$) 
trajectory in addition to the Pomeron which is not included for the ZEUS
measurement as it does not improve the quality of the fit in that case.
Corresponding value of the Pomeron intercept is $\alphapom(0)=1.173$
($\alphapom(0)= 1.132 \pm 0.006$) from H1 (ZEUS) fit. 
\\

As shown by H1 \cite{h1f2d97}, the Regge factorisation ansatz holds within 
the present precision of the measurement as the \fdtwof\ measurement is not
sensitive to the $\xpom$ value. The \qsq\ dependence 
exhibits a more important scaling violation than in the $F_2$
structure function measured in non-diffractive deep inelastic
scattering indicating that the exchanged object 
in diffraction has an important gluon content.
\\

\begin{figure}[htbp]
 %\vspace{-0.8cm}
 \vspace*{2.2cm}
 \begin{picture}(60,70) 
  \put(0,0){\epsfig{figure=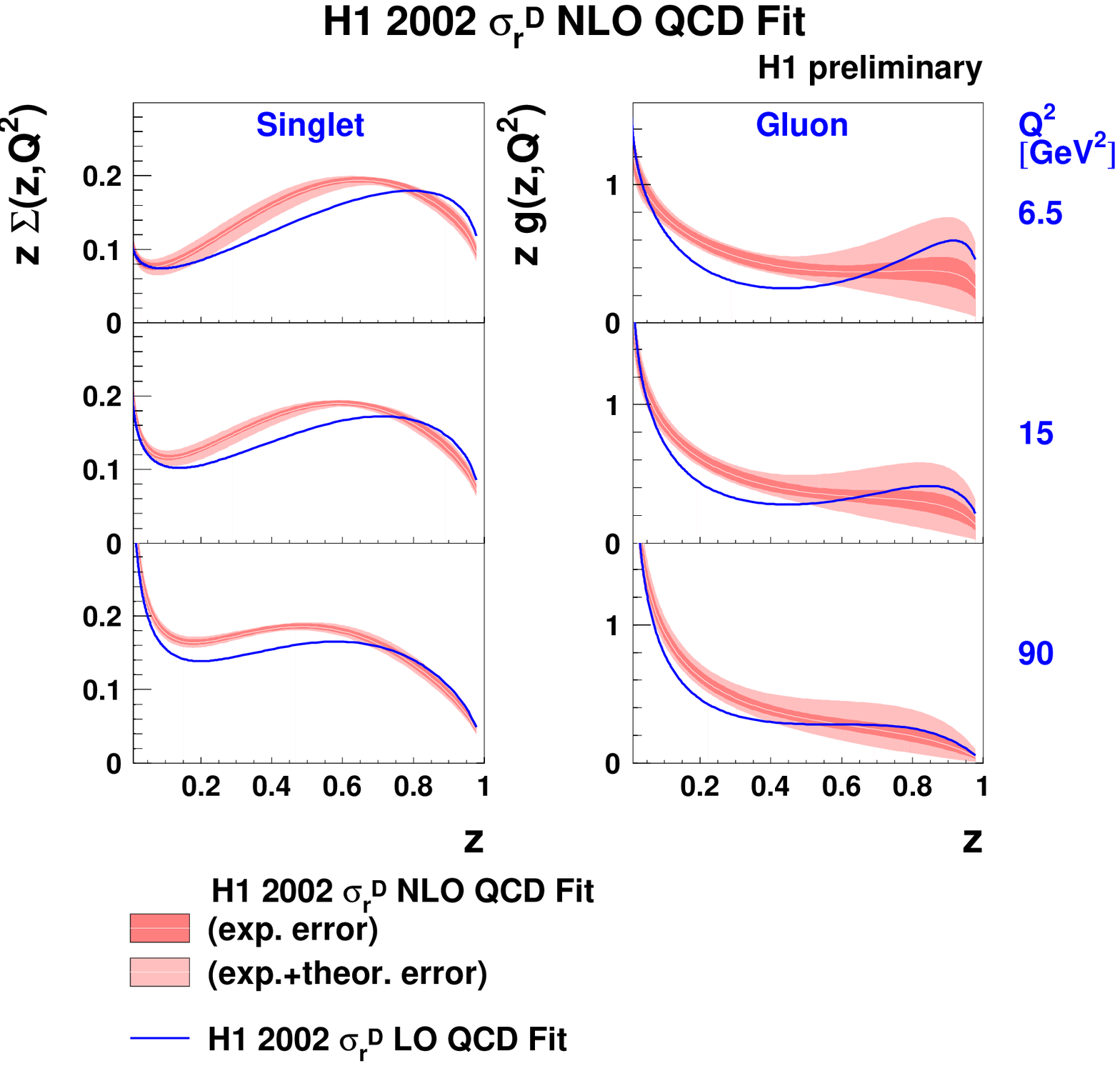,width=0.50\textwidth}}
   \put(167,5){\epsfig{figure=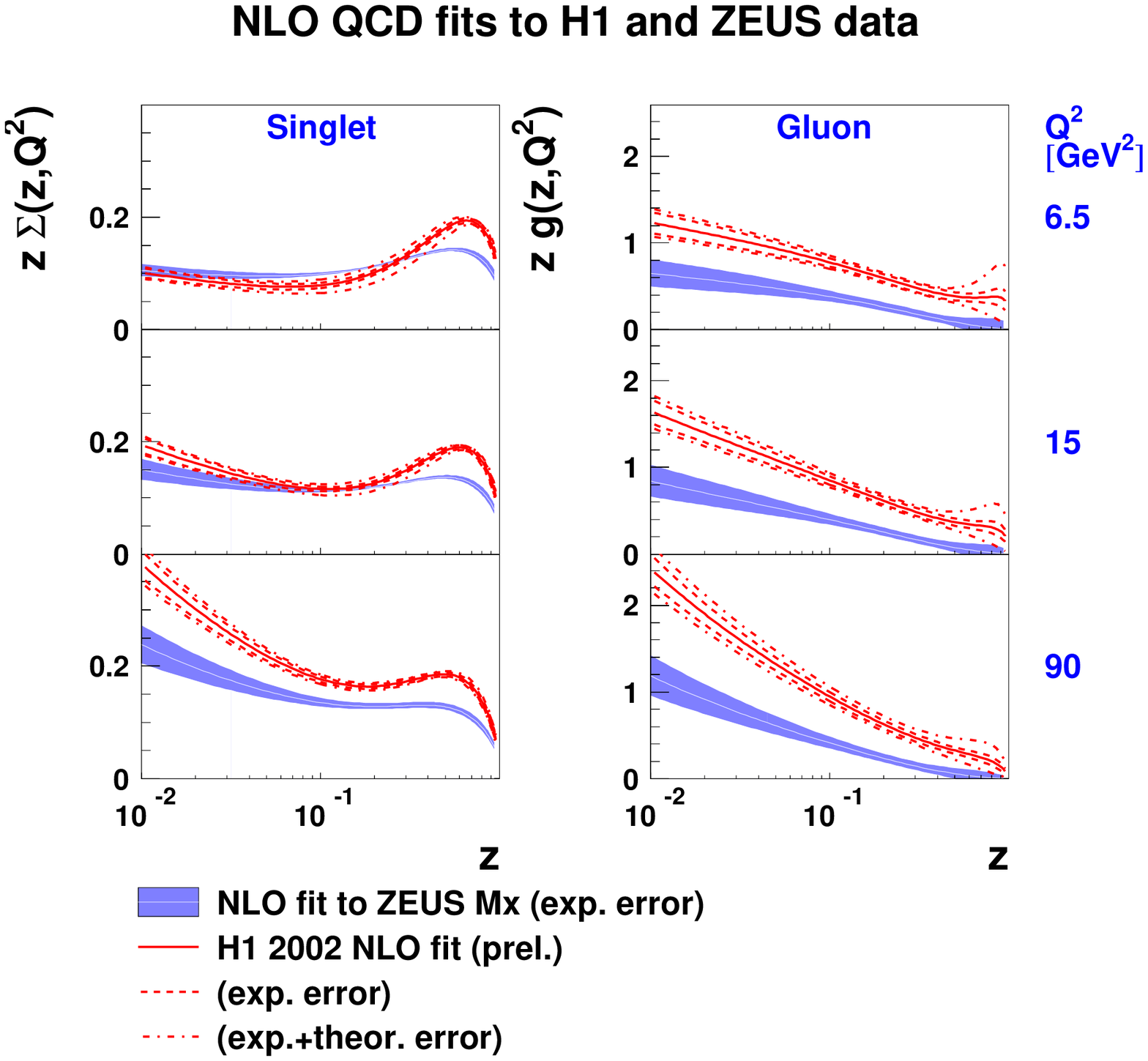,width=0.50\textwidth}}
 \end{picture}
%%\vspace{-0.5cm}
 \caption{Quark (singlet) and gluon densities in the $\pom$ extracted 
  from the QCD fit of \fdtwof\ as a function of $z$. {\bf left)} From H1
measurement. {\bf right)} From ZEUS measurement.}
\label{fig:gluond}
\end{figure}

 By analogy to the QCD evolution of the proton structure function $F_2$,
one can attempt to extract the partonic structure of the Pomeron
from the \qsq\ evolution of \fdtwo. Starting the QCD evolution at
$Q^2_0=6.5 \rm\ GeV^2$, extracted partonic 
distributions \cite{nlopaper,heralhc} are shown in Fig.~\ref{fig:gluond}
separately for the gluon
and the singlet quark components as a function of $z$, the Pomeron
momentum fraction carried by the parton entering the hard interaction.
The left 2 columns of the plots correspond to the singlet quark and the
gluon components extracted from the H1 measurement; the right 2 columns
from the ZEUS measurement.
%This distribution shows the dominance of gluons up to the high $z$
%values in the Pomeron partonic structure (75$\pm$15\% after
%integration in $z$). 
As a consequence of the differences between the H1 and 
ZEUS \fdtwo\ a weaker gluon component is found from the ZEUS fit.
\\

The QCD factorisation is expected to break down at large \bet\ values 
where higher twist terms may become important. A part of them corresponds 
to the 
contribution of the hard scale integration in the Pomeron
\cite{watt}.  Such a QCD Pomeron corresponds to the dominant term
for exclusive Vector meson contribution like $J/\Psi$ and for the 
Deeply Virtual Compton Scattering~\cite{favart}.
\\

\begin{figure}[htb] \unitlength 1mm
  \vspace{-0.8cm}
  \epsfig{figure=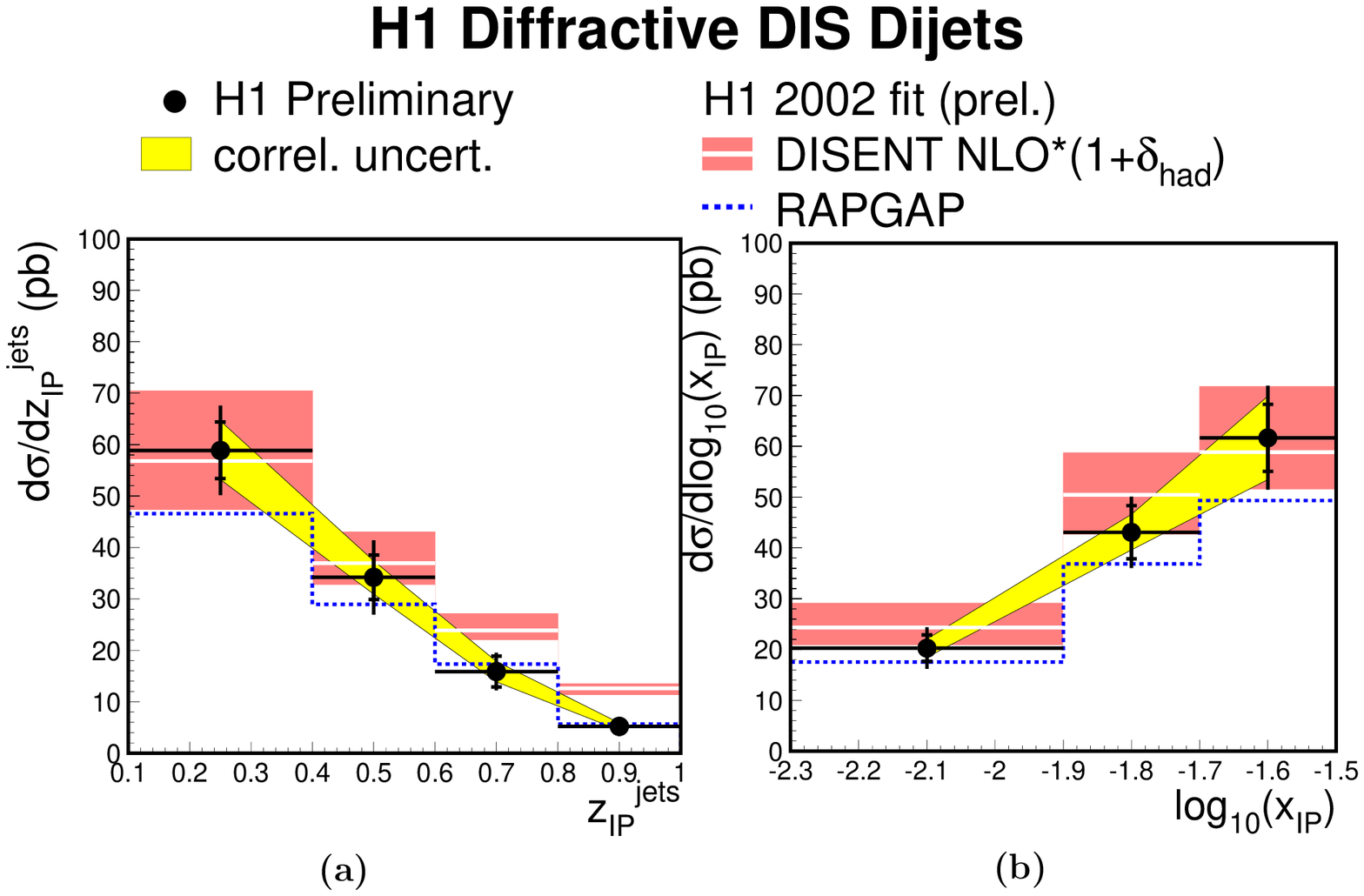,width=0.55\textwidth}
  \qquad
  \epsfig{figure=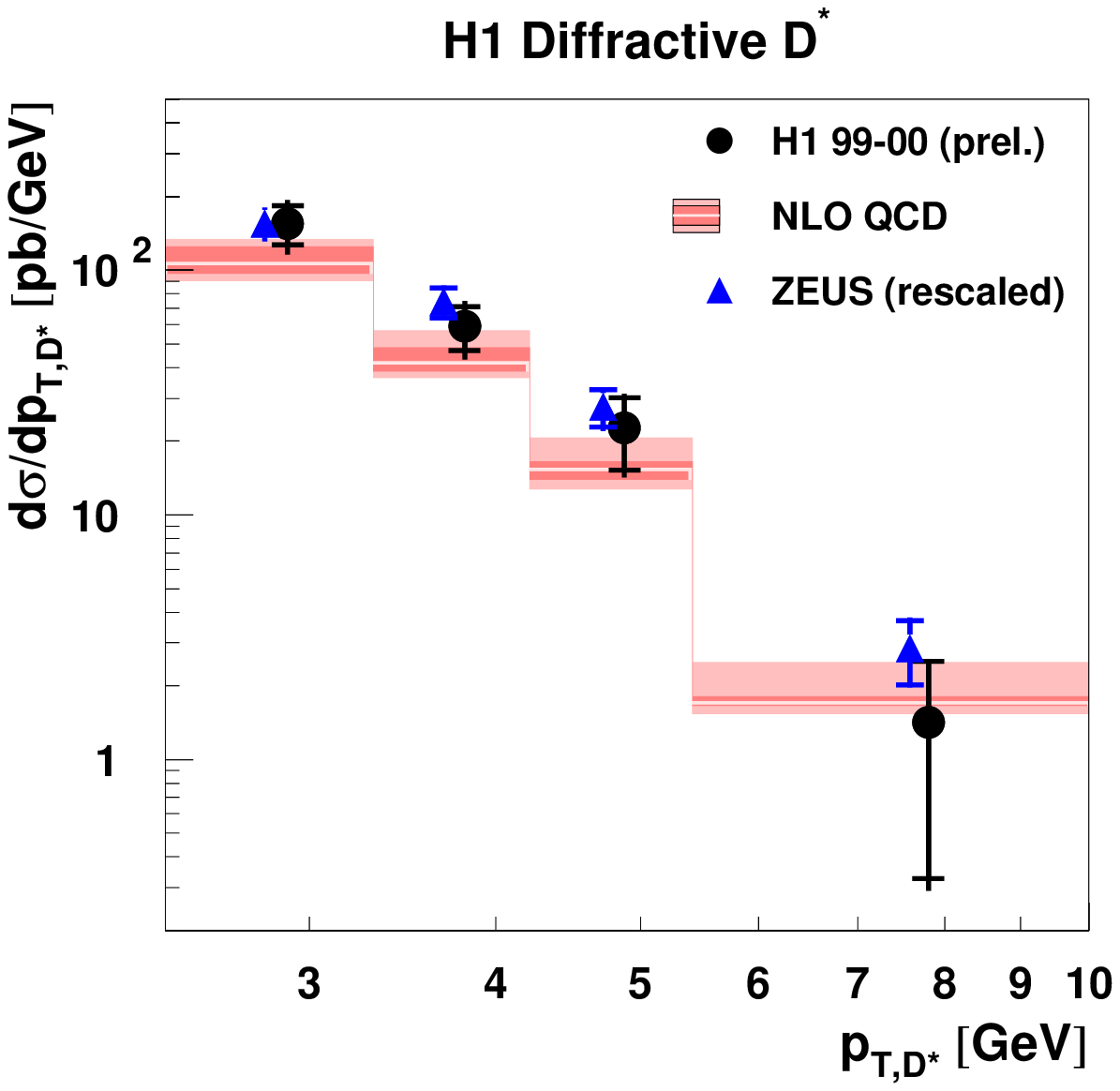,width=0.35\textwidth}
 \caption{{\bf left}: H1 measurement of diffractive dijet cross section 
 in electroproduction as a 
 function of $z_{\pom}^{jets}$, an estimator for the parton momentum fraction
 of the diffractive exchange entering the hard sub-process, and as a
 function of $\xpom$ as measured by H1 and ZEUS.
 {\bf right}: Diffractive $D^*$ meson cross sections in
electroproduction differential in $p^*_{T,D^*}$.
 }
\label{fig:dijet}
\end{figure}

{\bf Dijet and charm productions in diffractive electroproduction}\\
%%%%%%%%%%%%%%%%%%%%%%%%%%%%%%%%%%%%%%%%%%%%%%%%%%%%%%%%%%%%%%%%%%%
To test QCD factorisation for diffractive dijet production in
electroproduction regime ($Q^2 >> 1\rm\ GeV^2$), the
H1 dijet cross section~\cite{h1dijet} in the kinematic range 
$Q^2>4 \rm\ GeV^2$ and $\xpom<0.03$ is compared to the NLO
QCD prediction in Fig.~\ref{fig:dijet}, using the extracted 
diffractive parton densities as obtained by H1. The cross
sections were corrected to asymmetric cuts on the jet transverse
momentum $p_{T,1(2)}>5(4) \rm\ GeV$, to facilitate comparisons with
NLO calculations.
The inner error band of the NLO
calculations represents the renormalisation scale uncertainty, whereas
the outer band includes the uncertainty in the hadronisation
corrections. Within the uncertainties, the data are well described in
both shape and normalisation by the NLO calculations, in agreement with 
QCD factorisation. ZEUS measures the same process in an equivalent
kinematic range and comparing it to different diffractive parton
densities concludes: ``the differences observed between the 
sets of predictions may be interpreted as an estimate of the uncertainty 
associated with the dPDFs... 
A better understanding of the dPDFs and their uncertainties is required
before a firm statement about the validity of QCD factorisation can be made."
\cite{zeusdijet}.
\\

The  $D^*$ meson production measurements in
diffractive electroproduction were achieved by H1~\cite{h1charm} and
ZEUS~\cite{zeuscharm} for the kinematic range $Q^2>2 \rm\ GeV^2$,
$\xpom<0.03$  
and $p^*_{T,D^*}> 2 \rm\ GeV$, where the latter
variable corresponds to the transverse momentum of the $D^*$ meson in
the photon-proton centre-of-mass frame.
NLO QCD calculations were performed interfacing the H1 diffractive parton
distributions.
The renormalisation and factorisation scales were set to
$\mu^2=Q^2+4m_c^2$. 
%Further parameter values are a charm quark mass
%of $m_c=1.5 \rm\ GeV$, a $c\rightarrow D^*$ hadronisation fraction of
%$f(c\rightarrow D^*)=0.233$ and $\epsilon=0.078$ for the used Peterson
%fragmentation function.
A comparison of the calculations with the $D^*$ H1 and ZEUS data is shown in
Fig.~\ref{fig:dijet}-right. The inner error band of the NLO calculation
represents the renormalisation scale uncertainty, whereas the outer
error band includes variations of the charm mass and of the Peterson
fragmentation function. Within the
uncertainties, the data are well described in both shape and
normalisation by the NLO calculations, supporting the idea of QCD 
factorisation.

\section{Comparison with Hard Diffraction at the TEVATRON}

\begin{figure}[htb]
 \vspace*{-1.4cm}
 \epsfig{figure=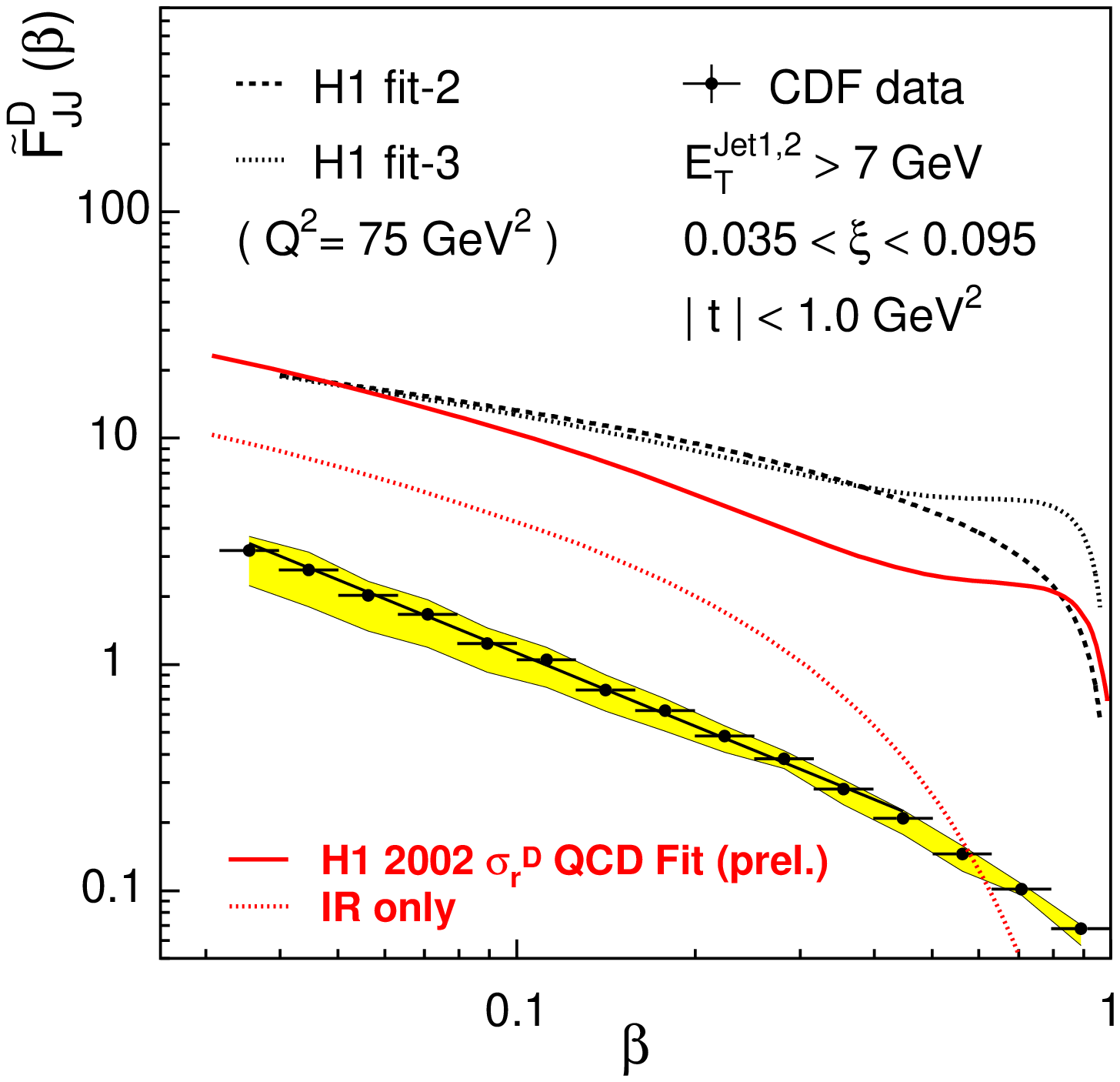,width=0.47\textwidth}
 \qquad
 \epsfig{figure=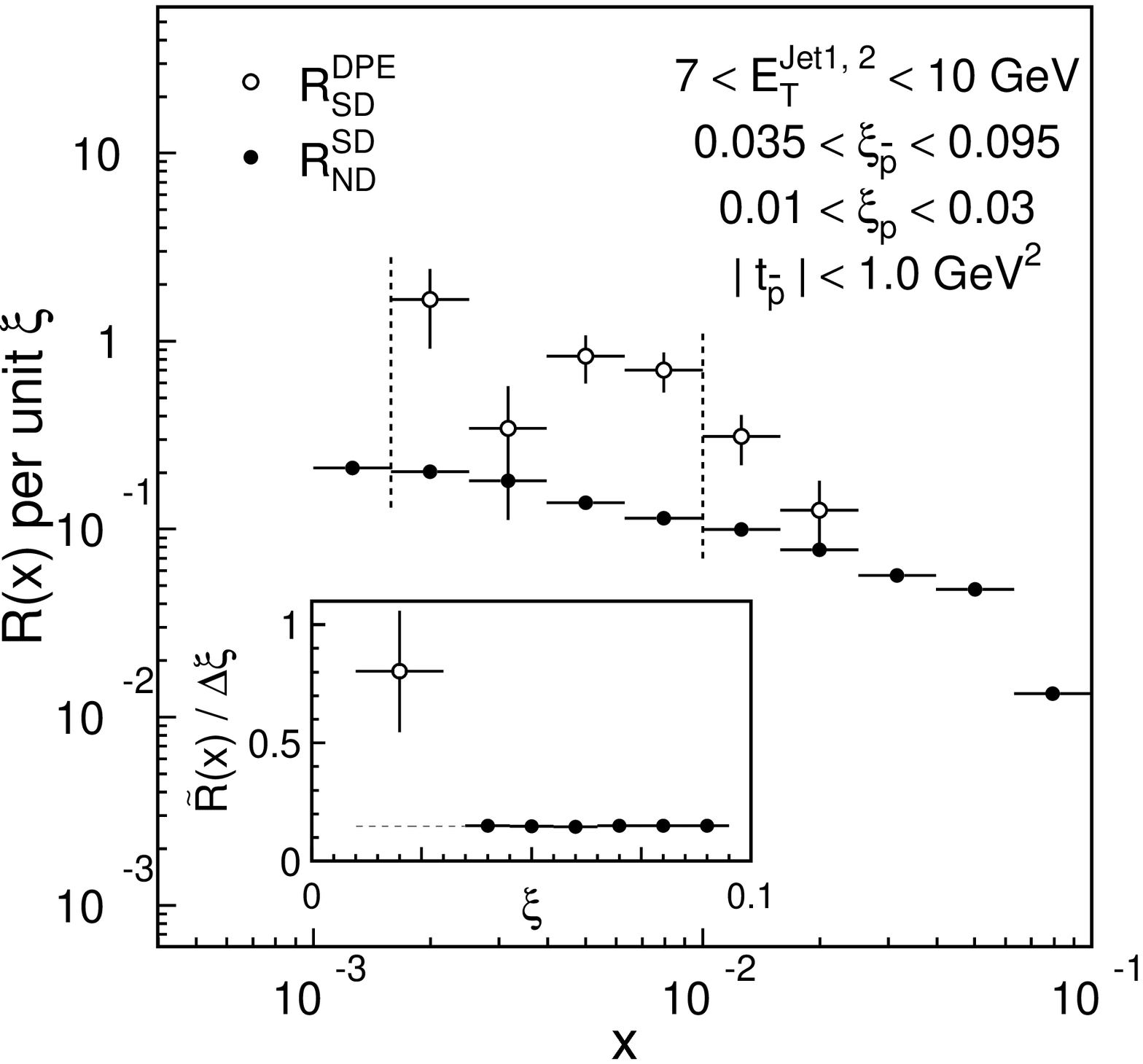,width=0.45\textwidth}
 %\vspace*{-1.0cm}
\caption[]{{\bf left }: measured dijet cross section in single-diffractive
processes in comparison with the prediction using the diffractive PDF's 
obtained by H1. {\bf right }:  measured dijet cross section in
single-diffractive processes compared to double-diffractive
processes by CDF. 
}
\label{fig:tevdijet-h1fit}
\end{figure}

One of the most striking features of the hard diffractive process
measured at the Tevatron is a large suppression of the cross section
with respect to the prediction based on the diffractive
parton densities obtained from the HERA $F_2^D$ data.
Figure~\ref{fig:tevdijet-h1fit}-left shows the comparison of the dijet
cross section in the single-diffractive process measured by CDF 
at the Tevatron~\cite{Tevatron-SD-dijet-1} to the
prediction using the diffractive parton densities discussed in
the previous section based on the H1 measurement. 
Although the prediction reproduces
the shape of the data in the low-$\beta$ region, the magnitude of the
cross section is smaller by a factor 5 to 10.
%A similar degree of suppression was observed in the other hard
%diffractive processes~\cite{Tevatron-SD-W,Tevatron-SD-dijet-2}.
This indicates a strong factorisation breaking
between HERA and the Tevatron: the diffractive parton densities
are not universal between these two environments.
Additionally, CDF has measured the ratio of double-diffractive over 
single-diffractive processes (shown in the right plot of
Fig.~\ref{fig:tevdijet-h1fit}) to be 0.19 $\pm$ 0.07. This indicates 
that the formation of a second gap is not (or only slightly) suppressed.  
\\

The reason for the breaking is not yet clearly known.
It is usually attributed to re-scattering between spectator partons in the 
two beam remnants where one or more colour-octet partons are exchanged,
which destroys the already formed colour-singlet state.
\\

{\bf Factorisation test with hard photoproduced diffraction at
HERA}\\
%%%%%%%%%%%%%%%%%%%%%%%%%%%%%%%%%%%%%%%%%%%%%%%%%%%%%%%%%%%%%
QCD factorisation can be further investigated within HERA
looking at the diffractive dijet photoproduction ($Q^2 \sim 0$),
where the hard scale is provided by the $E_T$ of the jets.
Factorisation is expected not to hold in photoproduction
events, where the resolved process has a photon remnant, allowing
re-scattering.
On the other hand, the direct process does not have
a beam remnant and the suppression of diffractive events is expected
to be much smaller than in the resolved process \cite{klasen}.
\\

\begin{figure}[tb]
 \vspace*{4.0cm}
 \begin{picture}(60,70) 
 \put(0,0){\epsfig{figure=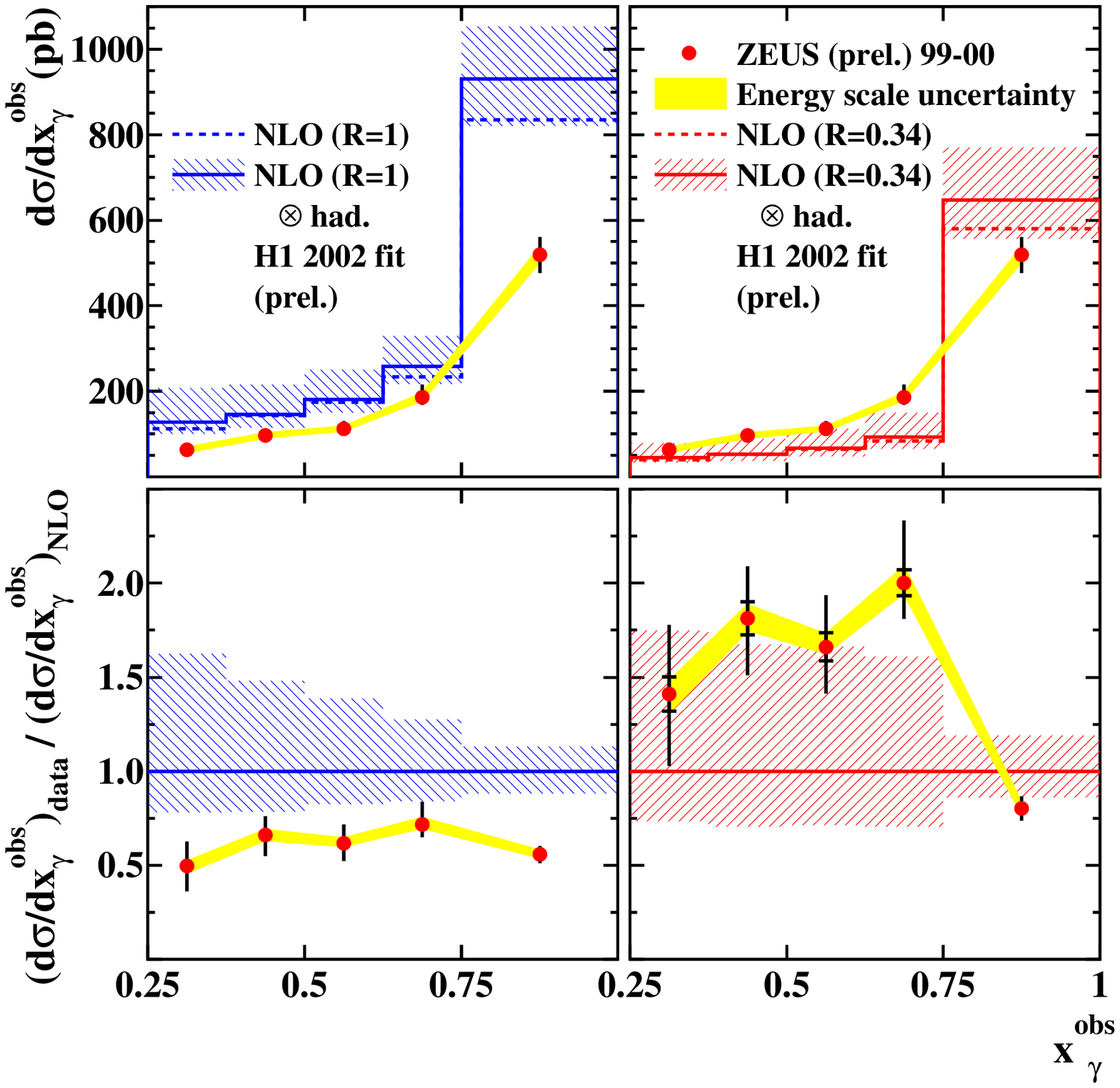,width=0.55\textwidth}}
 \put(220,40){\epsfig{figure=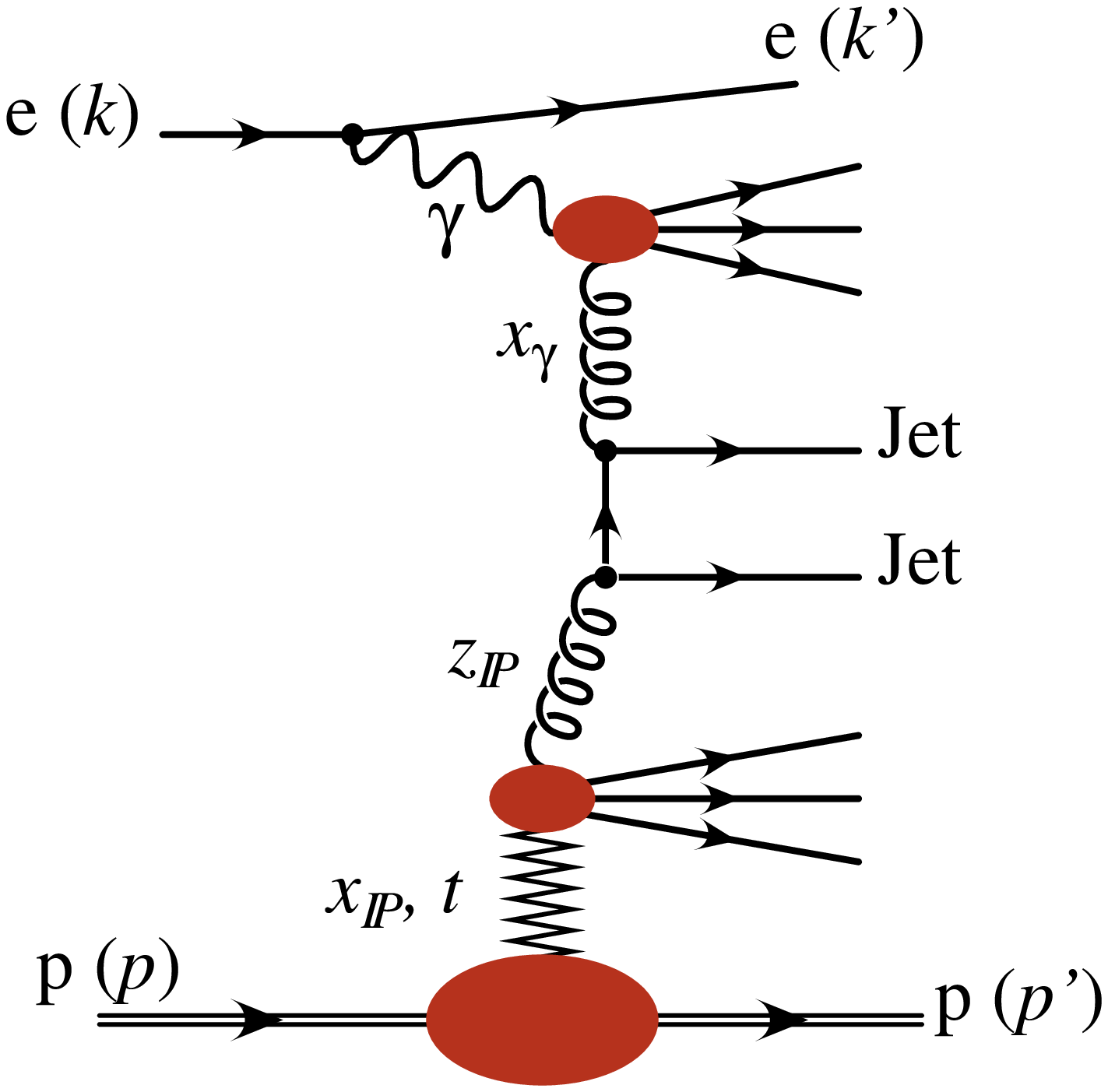,width=0.35\textwidth}}
 \end{picture}
\caption[]{{\bf left}: Dijet cross section in photoproduction at HERA 
measured by ZEUS as a function of $x_\gamma^{\rm jets}$. 
{\bf right}: Diagram for dijet in photoproduction at HERA.
}
\label{fig:phpdijet-xgam}
\end{figure}

Figure~\ref{fig:phpdijet-xgam}-left shows the dijet cross section in
diffractive photoproduction measured by ZEUS as a function of $x_\gamma^{jets}$
\cite{zeusdijet-gp},
the longitudinal momentum fraction of the parton that participated in
the hard scattering (see Fig.~\ref{fig:phpdijet-xgam}-right diagram), 
reconstructed from the dijet momenta.
Resolved events dominate in the low-$x_\gamma^{jets}$ region
while the direct process is concentrated at $x_\gamma^{jets}$ close
to one. 
The cross sections compared to the NLO QCD prediction exhibit a
factorisation breaking both in the direct and in the resolved parts. 
The NLO QCD prediction required a global factor 
of 0.5 to be able to describe the data.
Similar results have been measured by H1~\cite{h1dijet}.
\\

\begin{figure}
 \epsfig{figure=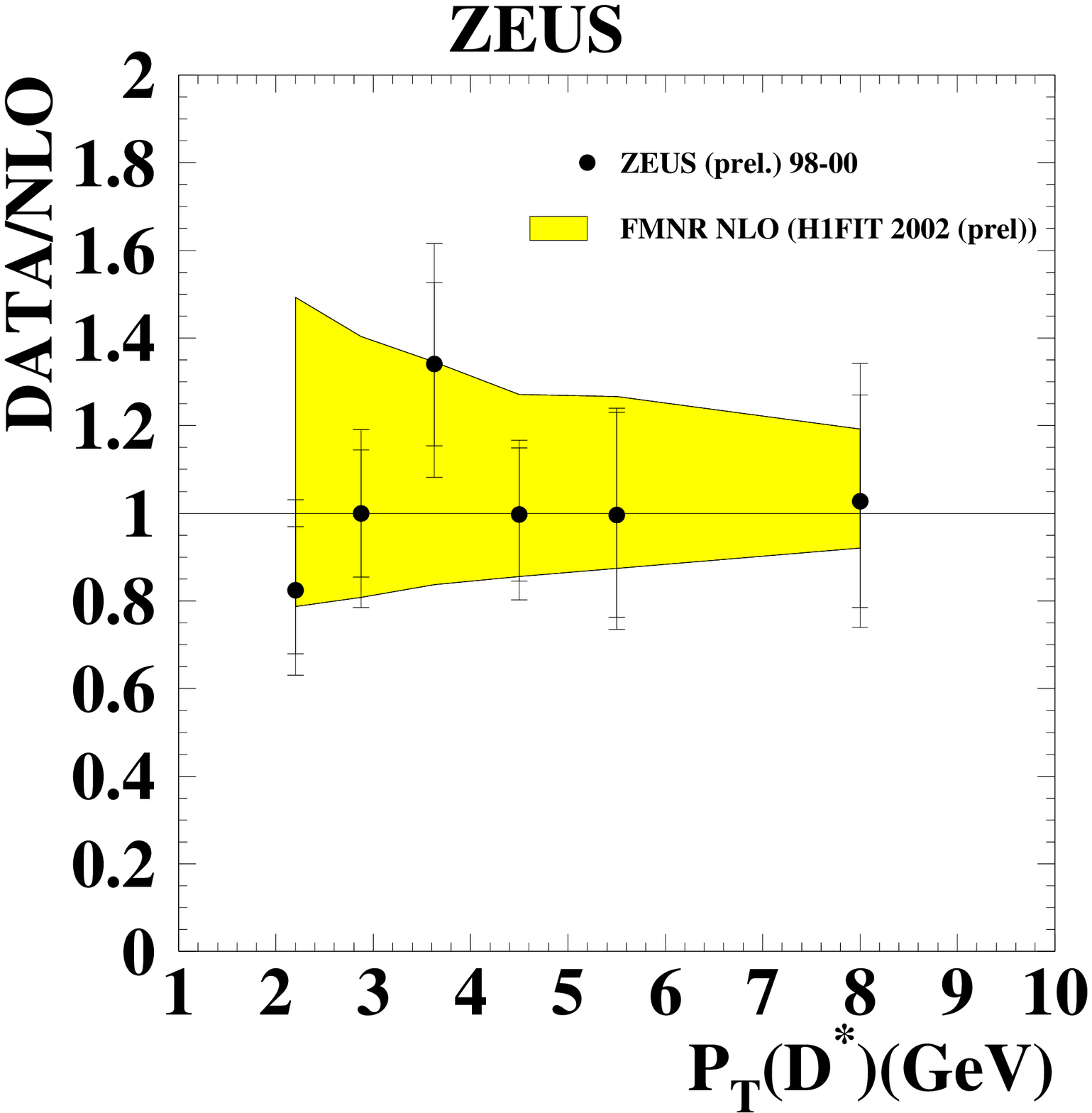,width=0.30\textwidth}
 \epsfig{figure=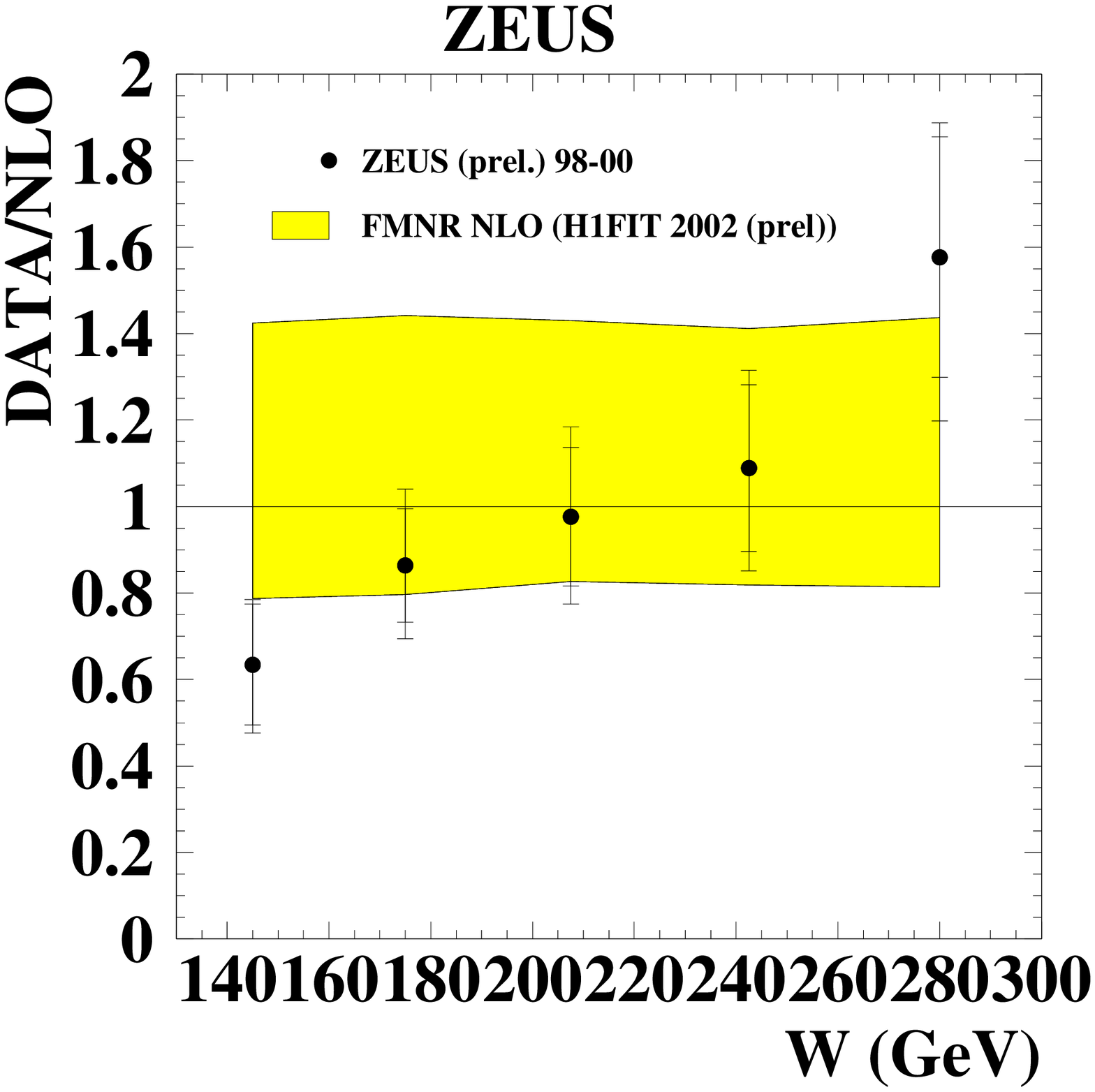,width=0.30\textwidth}
 \epsfig{figure=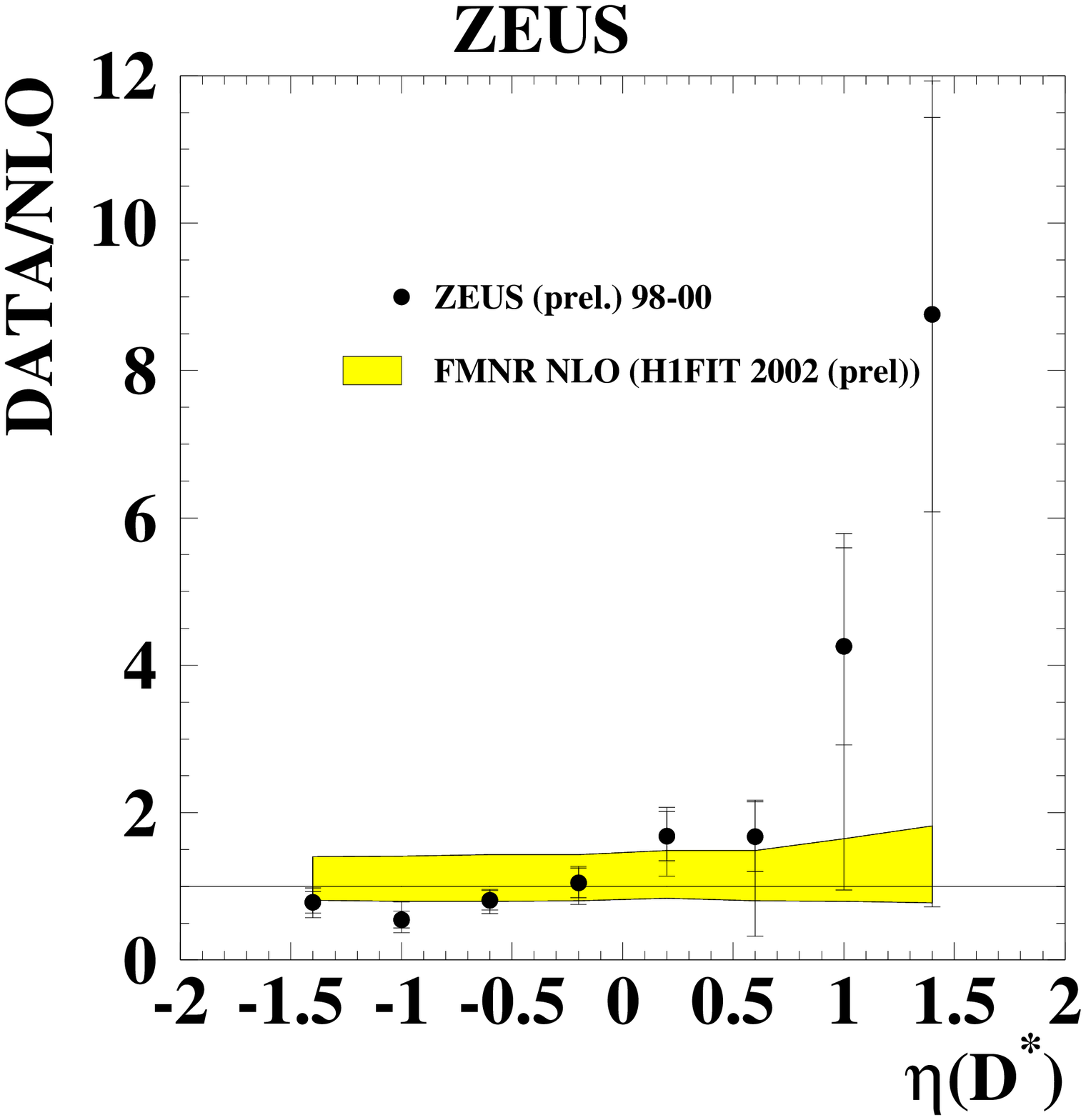,width=0.30\textwidth}
\caption[]{{\bf left}: Ratio of the $D^*$ meson cross section to the NLO
QCD prediction in photoproduction measured by ZEUS as a function of 
$P_T^{D^*}$, $W$ and $\eta$.
}
\label{fig:charmgp}
\end{figure}

Recently $D^*$ meson photoproduction cross section in
diffraction has been measured by ZEUS~\cite{zeuscharmgp} 
for the kinematic range 
$P_T^{D^*} > 1.9$ GeV, $\eta_{D^*}<1.6$, $ 130<W<300$ GeV
and $0.001 < \xpom <0.035$. 
The measurement is found to be in good agreement in shape and in
normalisation with the NLO QCD prediction as shown in Fig.\
\ref{fig:charmgp} presenting the data ratio to the theory.
% as a function of $P_T^{D^*}$, $W$ and $\eta$.

\section{Concluding Remarks} 
%%%%%%%%%%%%%%%%%%%%%%%%%%%%
After almost ten years of research at HERA 
we are reaching a perturbative QCD understanding of hard
diffraction in $ep$ interactions. 
The partonic structure of the diffractive exchange has been
measured and is found to be dominated by gluons. 
 In its validity domain ($Q^2 >> 1 \rm\ GeV^2$ in $ep$ collisions), the
QCD factorisation holds, as confirmed by charm and dijet productions.
In $p\bar{p}$ collisions, at Tevatron, a breakdown of the QCD factorisation
of a factor 5 to 10 is observed for one gap formation and no (or weak) 
additional gap suppression is observed for a second gap formation.
Rescattering corrections seem to be important in $p\bar{p}$ and in $ep$
collisions in dijet photoproduction. The $D^*$ photoproduction is in
agreement with the NLO QCD prediction within the present precision. 

A global
understanding of inclusive and exclusive hard diffractions has progressed
. 
Many more results and a deeper understanding are needed and expected with
the coming data at HERA II, Run II at Tevatron, Compass and in a further
future at LHC. 

\section{Acknowledgments} 
%%%%%%%%%%%%%%%%%%%%%%%%%%
It is a pleasure to thank the organizers of the Ringberg workshop on 
{\it New Trends in HERA Physics 2005} for their kind invitation and the 
perfect organization of this very interesting workshop. 
This work is supported by the Fonds National de la Recherche
Scientifique Belge (FNRS).

\end{document}